\documentclass[12pt,preprint]{emulateapj}

\def\gtorder{\mathrel{\raise.3ex\hbox{$>$}\mkern-14mu
             \lower0.6ex\hbox{$\sim$}}}
\def\ltorder{\mathrel{\raise.3ex\hbox{$<$}\mkern-14mu
             \lower0.6ex\hbox{$\sim$}}}

\shorttitle{Survival Rate of Ejected Earth-Moon Systems}
\shortauthors{Debes et al.}

\begin{document}
\title{The Survival Rate of Ejected Terrestrial Planets with Moons}
\author{John H. Debes\altaffilmark{1},Steinn Sigurdsson\altaffilmark{2}}

\altaffiltext{1}{Department of Terrestrial Magnetism, Carnegie Institution of 
Washington, Washington, DC 20015}
\altaffiltext{2}{Department of Astronomy \& Astrophysics, Pennsylvania State
University, University Park, PA 16802}
\vfill

\begin{abstract}

During planet formation, a gas giant will interact with smaller
protoplanets that stray within its sphere of gravitational influence. We 
investigate the outcome of interactions between gas giants and terrestrial-sized protoplanets with lunar-sized companions.  An interaction between a giant
planet
and a protoplanet binary may have one of several consequences, including
 the delivery of volatiles to the inner system, the capture of retrograde moons by the giant planet,
 and 
the ejection of one or both of the protoplanets.  We
show that an interesting fraction of terrestrial-sized planets with lunar sized
companions
will likely be ejected into interstellar space
with the companion bound to the planet.  The companion provides an additional source 
of heating for
the planet from tidal dissipation of orbital and
spin angular momentum.  This heat flux typically is larger than the current
radiogenic heating of the Earth for up to the first few hundred million years of evolution.  In combination with an atmosphere of sufficient thickness and composition, the
heating
can provide the conditions necesary for liquid water to persist on the 
surface of the terrestrial mass planet, making it a potential site
for life.
We also determine the possibility for
directly detecting such systems through all-sky infrared surveys or 
microlensing surveys.  Microlensing surveys in particular will directly measure
the frequency of this phenomenon.
\end{abstract}   


\section{Introduction}
Binary protoplanets of terrestrial mass and lower 
may be common--large Kuiper belt objects in particular
seem to have a high fraction of satellites that formed through giant impacts
much in the same way as the Earth/Moon system \citep{canup04, canup05}.
  Three of the four known
large KBOs (Pluto, 2003UB313, and 2003EL61) contain satellites, with Pluto/Charon having a mass ratio closest to 1 \citep{brown06}.  Assuming similar
 densities and albedos
between parent and body and based on the relative brightnesses, the other two binary KBOs
 have mass ratios similar to the Earth/Moon system \citep{brown06}.
Since these objects and the Earth/Moon system exist, three body interactions need to be studied in the context of planet formation.

Giant impacts, the main process for forming lunar mass moons to terrestrial planets, occur 1-2 times over the course of terrestrial planet formation simulations with enough angular momentum to form the Earth-Moon system \citep{agnor99,chambers01}.  The value of this angular momentum is dependent on the angle of impact, the mass of the impactor, and ratio of impact speed to escape 
speed.  A range of parameters can create a situation where
 Earth-Moon analogues
are formed.  This range is limited for the specific case of the Moon due to other constraints such as a lack of iron relative to the Earth \citep{jones89}, but many constraints are relaxed for an impact that creates a generic moon of lunar mass, allowing for a reasonable probability that the Earth/Moon system is mediocre and analogues should be present in a non-negligible fraction of planetary systems.

Even though there are examples of binarity in the terrestrial planet region
and in the Kuiper belt, it is not
clear whether conditions
exist to form Earth/Moon sized systems where giant planets arise.  Two scenarios could create a situation where a binary 
protoplanet may come in close contact with a Jovian planet.  

The first is that
the binary interacts with a Jovian {\em in situ} during giant planet formation.
The binary would be a leftover from
oligarghic growth, similar to the scenario
proposed by the formation of Uranus and Neptune through scattering with
Jupiter and Saturn \citep{thommes02}.  In some cases, oligarchic losers will beejected to escape speeds.    
The size distribution of protoplanets in this region is poorly known, but oligarchic growth starts at 
$\sim10^{-5}$ M$_{\oplus}$ and proceeds to a gas acrretion runaway
growth mass of $\sim$10 M$_\oplus$ \citep{thommes03,pollack96}.  Oligarchic growth
in the terrestrial planet region appears to spread from higher density to lower density regions \citep{ida02}, which suggests that a size range of oligarchs would be present in the region of 5-10~AU.  Leftovers that range
from the smallest oligarch mass to ice giant mass
are possible \citep{chiang07} and could explain Uranus's peculiar
axis of rotation as a result of a giant impact between it and a large
late stage leftover \citep{korycansky90,brunini02}.  The constraints on the size of left-overs at late stages of the Solar System's formation around the region of Neptune places
stringent constraints on the maximum size of the planetesimal distribution to sizes much smaller than the Earth, though one time encounters with Mars-like
protoplanets are not ruled out \citep{murray06}.  However, the typical architecture of a planetary system is not well known and the stochastic behavior of 
oligarchic growth would argue for a wide range of possible outcomes for
 planetary systems between 5-10~AU, including the presence
of terrestrial mass protoplanets \citep{levison98}.

Secondly, forming planetary systems where one or more giant planets migrate to small
orbital radii may lead to the ejection of several smaller planets
in the process \citep{mandell06}.  Some ejected
planets may have bound companions that are a significant fraction of their
host planet's mass.
The frequency of such systems and the 
survival rate of these binary protoplanets during ejection
is an interesting open question.

If terrestrial planets are ejected with lunar-sized companions, they may
 be habitable, despite being deprived of an insolating central
star.  If an isolated terrestrial planet is ejected with significant 
atmosphere captured from the parent protoplanetary disk, the planet will
have a surface temperature high enough to sustain 
liquid water \citep{stevenson99}. 

The presence of the companion
will add heat through tidal dissipation of spin and angular momentum, 
which will augment the heat flux due to radioactive materials in
the interior of the planet.  Tidal heating has been investigated in
the melting of the moon, the vulcanism on Io, and the heating of Europa \citep{peale78,peale79,cassen79,carr98}.   The heating will decay over time, but if life evolves, it could persist for long times below the surface of the larger planet \citep{laughlin00}.  

\section{Numerical Simulations}

\begin{figure}
\plotone{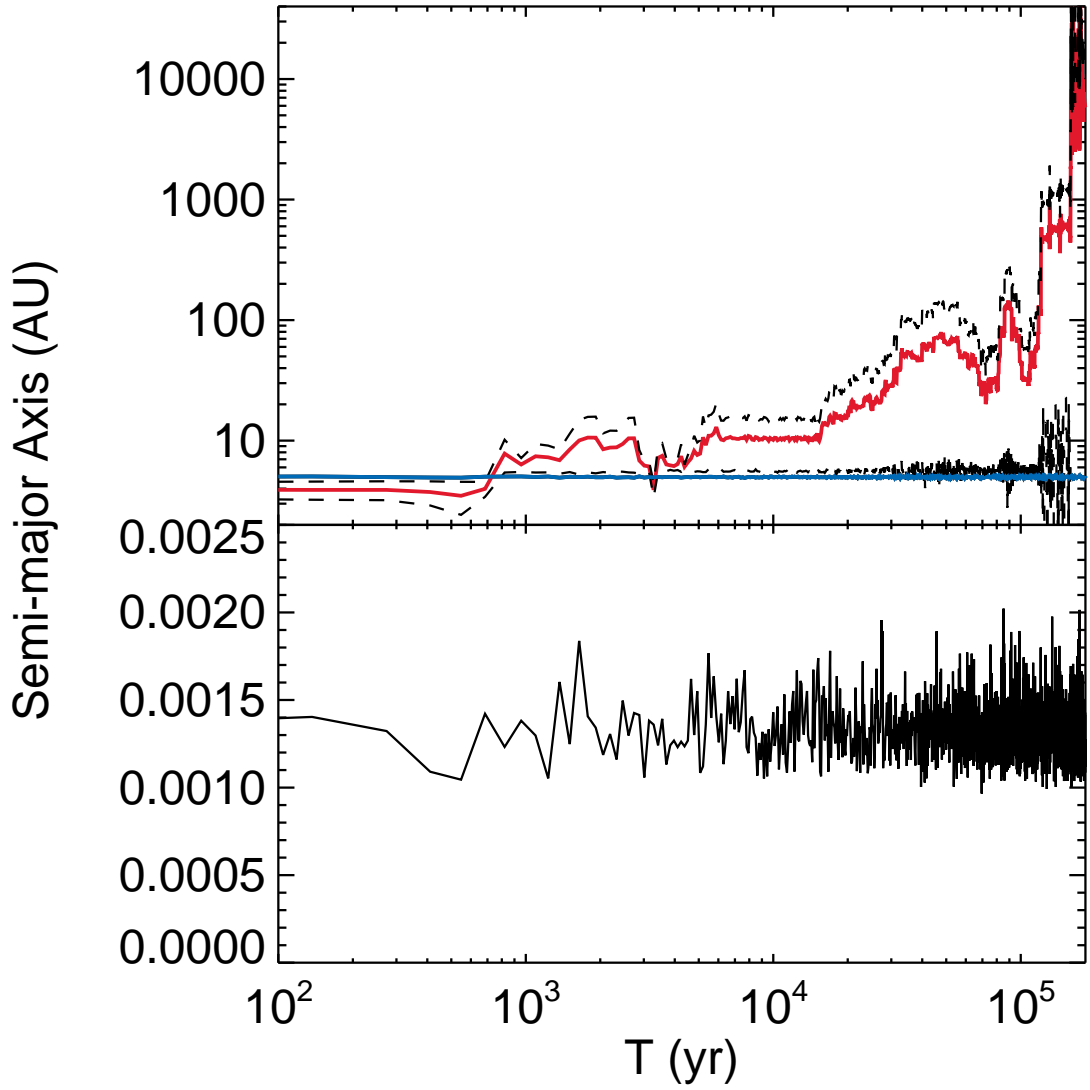}
\caption{\label{fig:avst}(Top) An integration of a binary protoplanet in a Hill unstable
orbit with a gas giant. The thick blue curve is the gas giant, while the dashed
red curve represents
 the binary protoplanets.  The thin dashed lines represent the periastron and apastron
of the binary pair.   (Bottom) The same integration but in the 
rest frame of the terrestrial mass member of the binary protoplanets, showing
the semi-major axis of its companion as a function of time.  The eccentricity
of the companion remains roughly constant at $\sim$0.15 }
\end{figure}

To test the frequency of binary protoplanet ejections we have run 2700 numerical simulations using the Mercury integration code\citep{chambers99}.  We placed
a binary protoplanetary system with an Earth mass primary and a 
lunar mass secondary in Hill unstable orbits relative to a giant planet on a concentric orbit. 
Hill stability is defined as a system that experiences no close approaches
between the consituent planets in the system \citep{hill,gladman93}.  
We chose initial positions for the terrestrial planet to be between
0.5-1 times the critical Hill radius interior and exterior to the giant planet.
Initial eccentricities
of the binary protoplanet
 system were set to 0.05 and inclinations were $<$ 5$^\circ$.
Initial semi-major axes of the companion's
 orbit were set to 30 R$_{\oplus}$,
half the current Moon-Earth separation.  Initial Moon-Earth separations would
be smaller, but we chose this separation as a compromise between simulation speed and physical reality, as this separation dictates the time step required to accurately integrate the equations of motion.
  Timesteps were set to $\sim$8 days and were adaptable.  A system was run for a maximum of 10$^6$\ yr but was checked at 10$^4$\ yr and 10$^yr$.
  In this way, highly
interacting systems that disrupted the binary protoplanets were removed 
early in the integration.  Our criteria for ejection was that
the binary protoplanets were bound to each other, but unbound from the central
star with a radial distance $>$ 15~AU from the central star.  In the 
simulations, no system was considered ejected unless it passed beyond a radial
distance of 5000~AU, so in most cases the ejected binary protoplanets were 
allowed to evolve for several thousand years beyond the perturbation
that ejected the system.
Fractional errors in energy and angular momentum were $\sim$10$^{-11}$ but were
 larger if systems suffered a collision.  The highest errors
were $\sim$10$^{-8}$.  Figure \ref{fig:avst} shows a typical ejection of a 
protoplanet with a surviving companion.

As a further test of the stability of the simulations we
 ran ten long term simulations of an isolated binary protoplanet system to 
determine if the initial orbit could be stable for 10$^6$ yr, the maximum
length of the simulation.  We found that in all ten simulations the orbits 
remained stable for initial eccentricities of $\sim$0.15.

\section{Results}
 
Of the 2700 simulations, 90, or 3.3\% of the interactions resulted in only the
Earth being ejected,
while 123 or 4.6\% of the simulations ended in an ejection of a bound Earth-Moon system.
In most cases the eccentricity of the bound Earth/Moon systems
increased after ejection.  Bound ejections occur through
multiple grazing encounters with the Jovian, which is consistent with the
perturbations on the Earth/Moon system being less than their binding energy.

An estimation of the tidal dissipation in these ejected systems
shows that tidal heating can be as important as radioactive heating.
Heating can come from circularization of a satellite's eccentric orbit as
well as synchronization between the planet's spin and the satellite's mean
motion.  Heating rates for circularization are dependent on the eccentricity $e$, semi-major axis $a$, and mean motion $n$:
\begin{equation}
 \dot{E}_{circ}=-\frac{63e^2n}{4\tilde{\mu}Q_p}\left(\frac{R_p}{a}\right)^5\frac{Gm^2_s}{a}.
\end{equation}
The variable $\tilde{\mu}$ is the ratio of elastic to gravitational forces, or the effective rigidity of the planet. The variable $Q_p$ is the planet's specific dissipation function, and R$_p$ is 
planetary radius.  Heating for synchonization can be found by the following
equation:
\begin{equation}
\label{eq:synch}
\dot{E}_{synch}=-sign(\omega-n)\frac{3k_{2p}}{Q_p}\frac{m_s^2}{m_s+m_p}\left(\frac{R_p}{a}\right)^5n^2a^2(\omega-n)
\end{equation}
where we assume that the planet's
rotational frequency is greater than its mean motion and the rotational period
for the terrestrial planet is 10 hr \citep{mcdermott}.  Median values of
eccentricity and semi-major axis of the ejected pairs are 0.13 and 30 R$_\oplus$, respectively.

We assume two cases, a rocky planet and an icy planet.  A 
protoplanet will fall in between these two cases, depending on where it formed.
A rocky planet will be identical to the Earth in mass, radius, density,
and rigidity.  We take $k_2$=0.299, $\tilde{\mu}$=4 and Q$_p$=12 \citep{mcdermott}.   For an icy body we assume a rigidity of ice of 4$\times$$^{12}$\ dyne cm$^{-2}$, and assuming a density of 1 g cm$^{-3}$ for pure water ice, a $k_{2p}$=0.7 and $\tilde{\mu}$=1.  We assume a $Q_p$=100. 

\begin{figure}
\plotone{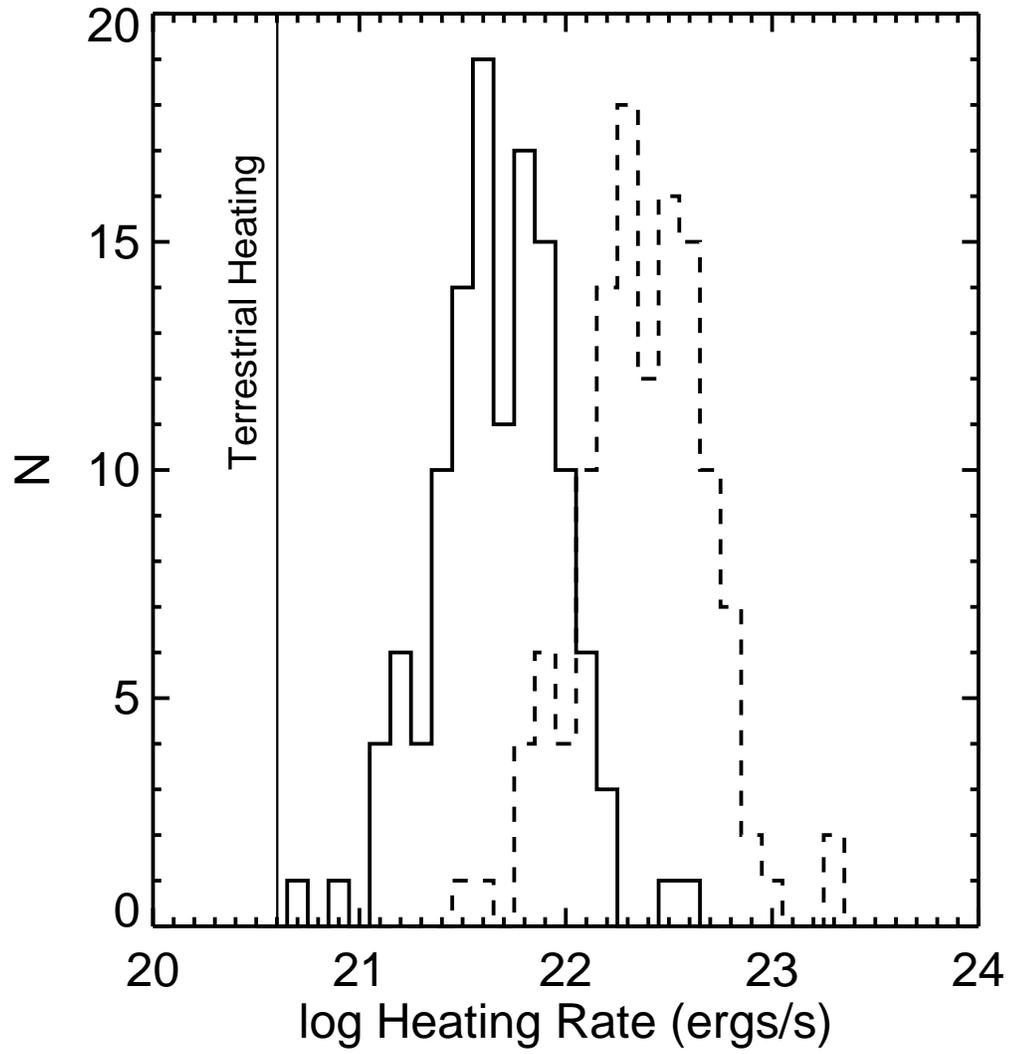}
\caption{\label{fig:hist}The distribution of tidal heating for the 123 ejected
binary protoplanets.  The solid line curve 
corresponds to the assumption of rocky
planets, while the dashed line corresponds to the assumption of icy planets.
The vertical line is the amount of radiogenic heating currently on Earth.}
\end{figure}

From among the end states of the ejected planets, the maximum heating came from a system with a semi-major axis of 21 R$_\oplus$
and an eccentricity of 0.21, corresponding to a total heating of
4.2$\times$10$^{22}$ ergs/s assuming a rocky planet.  This is 100 times larger than current terrestrial
radiogenic heating\citep{stacey92}.  The maximum heating for an icy planet is 
a factor of five larger.  The distributions of heating rates for
rocky and icy planets from our simulations relative to the
terrestrial radiogenic rate are presented in Figures \ref{fig:hist}.

 We compare these rates to the tidal heating on 
Io, whose luminosity is $\sim$10$^{21}$ ergs/s \citep{veeder94},
or about three times larger than the terrestrial radiogenic heating rate.
The highest heating of a rocky planet 
is roughly 4\% of the total insolation the young Earth
 experienced from the Sun, assuming a current rate of 1.5 $\times$10$^{24}$ ergs
 s$^{-1}$.  While this heating is small in absolute terms compared to the 
conditions present on Earth for the formation of life, it does not create
an insurmountable barrier for life to form and evolve.
The heating would provide energy for maintaining
liquid water and forming life on the planet, potentially in localized
hot spots where vulcanism or other geothermal processes occur. 

Heating decay
over time for the terrestrial planet would depend on the change in the planet's
rate of rotation ($\dot{\omega}$) and the semi-major axis ($\dot{a}$) due to
tidal evolution of the pair.  Assuming $\dot{a}\propto a^{-5.5}$ and $\dot{\omega}\propto a^{-6}$\ \citep{mcdermott}, we calculated $a$ and $\omega$ as a function of time and determined
when $\dot{E}$ in Equation \ref{eq:synch} meets current
radiogenic levels.  For our best case, we found that the heating from Equation \ref{eq:synch} reaches radiogenic heating in 144 Myr for a rocky planet, and
246 Myr for an icy planet.  This may be enough time
for life to arise and adapt to decreasing temperatures on the
planet.

\section{Detectability of Free Floating Earths}
Given the uncertainties in the formation rate of binary protoplanets,
the number of such objects per star, and the number of planetary systems, we
 rely on an analogue of the Drake Equation to estimate the space 
density of ejected terrestrial 
planets and ejected binary protoplanets.  In reality there
will be a size distribution of such objects each with different rates, but this
will allow us to get an idea of detectability through imaging
and microlensing surveys.

\begin{equation}
N=N_{\star}f_{planets}n_{planets}f_{binary}f_{ejected}
\end{equation}
where N$_\star$ is the number of stars in the Galaxy, f$_{planets}$ is the frequency of planetary systems around stars, n$_{planets}$ is the number of
terrestrial sized protoplanets that are formed in the system, 
f$_{binary}$ is the frequency of
protoplanets that have companions, and 
f$_{ejected}$ is the frequency these protoplanets are ejected with their
companions intact.  With the exception of N$_\star$ and f$_{ejected}$, these quantities are not
known, but can be estimated within an order of magnitude; so a reasonable
estimate is possible.

We assume that the fraction of stars that have planets is 50\% as a compromise
between the minimum of a few percent, which would be the fraction of solar type stars with known planets, and a maximum of one planetary system per star. We assume
that the number of terrestrial 
protoplanets available for ejection is on the order of 10,
which is the same order of magnitude of rocky material present in the giant
planet region.  We assume that there are several contenders to 
become the eventual large giant planet, and the ``losers'' form the primary
resevoir of unstable protoplanets that will encounter the Jovian.  Systems with Jovians that migrate will also play a role, which 
may help to increase this resevoir.

The fraction of protoplanets that have companions is also unknown, but can 
be assumed to be roughly 33\% based on our own Earth-Moon system
 and the assumption
that our Solar System is mediocre.  Considering the frequency of binary 
asteroids and kuiper belt objects, the frequency is probably no less than a 
few percent.

With these assumptions, we arrive at a total number of free floating
binary planets of 7$\times$10$^8$, assuming a total of 10$^{10}$ stars in the galaxy.
  Assuming that these planets are restricted to the
galactic thin disk, with a radius of 25~kpc and a scale height of 0.4~kpc, 
we estimate a space density for these objects of 9$\times$10$^{-4}$/pc$^3$.

To estimate the free floating planets' spectrum, we assume a median heating luminosity of 5.3$\times$10$^{21}$ ergs/s for rocky planets and a Planck spectrum.The flux density will peak at $\sim$77\micron, so
one would require an all-sky IR survey sensitive to $\sim$0.3  nano-Jy
 to detect
$\sim$100 objects out to a distance of 50~pc.  The background limited sensitivity of Hershel's PACS instrument is predicted to detect 3mJy sources at $\sim$68\micron.  A similar sensitivity would require a $>$ 750m size primary to detect a free floating planet in 1 hour of observing.  However, the closest of these
objects will be 5 times closer and thus 25 times brighter with a large proper
motion.  Provided it can be resolved from the IR background due to galaxies,
one to a few detections could be made with less stringent restrictions.  Thus,
we consider the above to be a conservative estimate of what could be
detected. 

In the case of microlensing, proposed next generation surveys
will be sensitive to free floating terrestrial planets \citep{rhie02}.  In 
most cases, microlensing events of planets in orbit around stars have a 
degeneracy between mass, distance, and proper motion.  This degeneracy is fortunately
broken given certain observational setups which allow
the other parameters to be measured \citep{han04,han05}.  
Binary protoplanets with separations similar to those studied
in this work should be easily detectable given that the orbits reside well
within the Einstein ring radius of the host planet \citep{rhie02}.  Simulations
of detection frequencies for next generation missions assuming a free floating planet frequency of 1-10 free floaters per star result
in $\sim$10-100 detections of earth mass free floating planets for 
the mission \citep{rhie02}.  Assuming a third of free floating planets started out with companions and $\sim$5\% of those were ejected with a companion intact, there would be a 1-2\% chance that any free floater would have a lunar mass companion.

\section{Other Implications}
There are other implications
 that deserve further study.  A significant fraction of stable binary
systems at the end
of our simulations were at large radii ($\sim$1\%), but had not been ejected.  They will either continue to be on wide orbits, or they
will eventually be ejected on timescales longer than 1~Myr.  If some of these
systems remain bound to the central star, they may become rogue planets that
periodically perturb planetesimals at larger orbital separations, akin to the
``Planet X'' hypothesis presented to explain periodic extinctions on the Earth
\citep{whitmire85}.  For binary protoplanets, this occurrence appears to be rare.
It will be more common for singular protoplanets, but we did not follow this specific outcome.

Of the 2700 simulations 12\% of the lunar companions were captured into resonant retrograde orbits with the Jovian planet.  Tidal dissipation between the 
companion and the Jovian could be enough to capture the companion into a 
stable bound orbit.  Three body interactions have been
pointed to as a formation mechanism for the retrograde Neptunian 
satellite Triton, and
could explain how the Saturnian satellite Phoebe could have been captured from
the outer solar system into its current elliptical and retrograde orbit
\citep{johnson05,agnor06}.

Several simulations had remnant protoplanets that were kicked into eccentric
orbits that strayed into the inner system, with periastrons that approached
1.5~AU, thus providing a mechanism to bring volatile rich protoplanets into
the inner system.  The delivery of volatiles from further out in the 
proto-planetary disk seems to
be a natural outcome of planet formation \citep{morbidelli00,raymond04}.
  
Tidal heating is greater around planets that are more massive, especially terrestrial companions to gas giants or brown dwarfs.  These objects will either be large oligarchic losers or isolated objects that formed with large circumstellar disks \citep[i.e.][]{luhman06}.  The heating these objects feel
will be more substantial than that felt for the planets in this
study and could be longer lasting if multiple companions are formed whose 
mutual gravitational interactions provide eccentricity perturbations.  Each
of these larger planets will have respective habitable zones where tidal
heating replaces insolation as the dominant heat source for
liquid water\citep{reynolds87,scharf06}.  These 
objects will also be detectable through microlensing.

Since all of the ejected binary protoplanets experience heating greater
than that currently experienced by the Earth, they will require thinner atmospheres than those considered by \citet{stevenson99}.  From
\citet{stevenson99}, the surface
temperature of a protoplanet with an atmosphere is 
\begin{equation}
T_s\sim425\chi^{\frac{1}{12}}\left(\frac{f_{atm}}{0.001}\right)^{0.36}
\end{equation}
where $\chi$ is the ratio of tidal heating to the current radiogenic heating
of the Earth, $f_{atm}$ is the fraction of the planetary mass that is 
in the form of an atmosphere.  To maintain a surface temperature of 270~K
with radiogenic and median tidal heating together requires two times less
 atmosphere
by mass than radiogenic heating alone.  Since tidal heating decays rapidly, radiogenic heating will be an important mechanism for sustaining an established biosphere for longer time scales.

\acknowledgements
The authors thank J. Chambers for helpful discussions on oligarchic growth and the referee, D. Stevenson for suggestions on improving the paper.

\clearpage

\end{document}